\documentstyle[12pt]{article}
\begin{document}

\title{Frequency Switching of Quantum Harmonic\\
Oscillator with time-dependent frequency}
\author{A. Angelow\\
Institute of Solid State Physics, 72 Trackia Blvd., Sofia 1784, Bulgaria\\
E-mail: Angelow@bgcict.acad.bg\\}

\maketitle

\begin{abstract}

An explicit solution of the equation for the classical harmonic
oscillator with smooth switching of the frequency has been found .
A detailed analysis of a quantum harmonic oscillator with such frequency
has been done on the base of the method of linear invariants. It has
been shown that such oscillator possesses cofluctuant states, different
from widely studied Glauber's coherent and "ideal" squeezed states.

\end{abstract}

Oscillator models are widely used in many branches of physics, such as
quantum optics, atomic, molecular, and solid state physics. Small
vibration of dynamic system can be describe in terms of harmonic
oscillators in both quantum and classical mechanics. To include
surrounding influences on the vibration, or to simulate the coupling
of the vibration with other degree of freedom, one can consider
time-dependent parameters specifying the Hamiltonian of a
harmonic oscillator, for example, mass and frequency. Besides,
nonstationary oscillator models show essentially nonclassical effects,
such as squeezing and covariance of their quantum fluctuations.
Some examples for such phenomena are: motion of one ion in Paul trap,
which is precisely described by harmonic oscillator, with periodically
time-dependent frequency \cite{paul}, also Berry phase can be
achieved when parameters of oscillator undergoes a cyclic change
\cite{moralles}-\cite{ditrich}. Agarwal and Kumar have shown that a
nonstationary oscillator with linear sweep of the restoring force
owns nonclassical states \cite{agarwal}. In the present Letter
we study the switching of the frequency of a quantum nonstationary
oscillator by using the method suggested in
\cite{trifonov3,trifonov4}.\ \

The Hamiltonian of a harmonic oscillator is given by
\begin{equation}\label{eq:HH}
{\hat H}= {1\over 2 m}{\hat p}^2 + {m \Omega^2(t)\over 2}{\hat q}^2 ,
\end{equation}
where the constants $m$ and $\Omega(t)$ are mass and the frequency of the
quantum harmonic oscillator. The case $M=M(t)$ can be reduced to the case
$M=m$ \, ( see for example \cite{trifonov3}, eq. 122 ).\ \

We recall the method of linear invariants, developed in
series of papers \cite{trifonov1}-\cite{trifonov4}, which
we apply to one dimensional case:
For each {\it quantum} system, described by a quadratic Hamiltonian,
there is a {\it classical} two- dimensional isotropic nonstationary
harmonic oscillator  with a Lagrangian ({\it classical})
\begin{equation}\label{eq:lagrangian}
{L={{m\over2}(\dot{\epsilon}_1^2+\dot{\epsilon}_2^2) -
{m\over2}{\Omega^2(t)}(\epsilon_1^2+\epsilon_2^2)}} .
\end{equation}
and equations of motion
\begin{equation}\label{eq:lag_equ}
{d\over dt} {\partial L\over \partial {\dot{\epsilon}}_k} -
{\partial L\over \partial \epsilon_k}=0 , k=1,2 .
\end{equation}
As it is shown in \cite{angelow}, these two real equations
are equivalent to  one complex classical equation of nonstationary
harmonic oscillator
\begin{equation}\label{eq:sys_equ}
{\ddot \epsilon(t)} + \Omega^2 (t)\epsilon(t)=0 ,
\end{equation}
where complex function $\epsilon(t)=\epsilon_1(t)+i\epsilon_2(t)$
completely describes the quantum evolution of the system
\cite{trifonov4}, in particular with Hamiltonian (\ref{eq:HH}).\ \

Despite of various investigations of nonstationary harmonic oscillator,
smooth switching of the frequency for finite interval of time,
was not presented in the literature. Here we study
the behaviour of the nonstationary harmonic oscillator
with varying frequency and constant mass for finite interval of time.
We will show that the found {\it classical} solution for this case
completely determines the {\it quantum} evolution of the corresponding
quantum oscillator too.\ \

Let we consider switching of the frequency $\Omega(t)$ in the form;
\begin{equation}\label{eq:Omega}
\Omega(\,{t}\,) = \left \{
\begin{array}{lr}
{ \omega}\,\sqrt {1 - {\displaystyle \frac {{
 \alpha}\,{ \omega}}{ \left( \! \,1 + { \alpha}\,
{ \omega}\, \!  \right) ^{2}}}} = \Omega_0 & \infty < t < 0\\
{ \omega}\,\sqrt {1 - {\displaystyle \frac {{
\alpha}\,{\omega}}{\left(\!\,1+{\alpha}\,{\omega}\,cos
(\,{\omega}\,{t}\,)^{2}\,\!\right)^{2}}}} & 0 \leq t\leq{\pi \over 2\omega}\\
{ \omega}\,\sqrt {1 - \alpha\,\omega }  & {\pi \over 2\omega} < t < \infty,
\end{array} \right.
\end{equation}
where $\alpha$ and $\omega$ are real constants,
with dimension of time and frequency, respectively. The shape
of the frequency is shown on figure 1. By direct calculation we can check
that the function
\begin{equation}\label{eq:epsilon}
\epsilon(t)= \left \{
\begin{array}{lr}
\sqrt {{{1+\alpha\,\omega}\over\omega}} \cos(
\omega\sqrt{1 - {\frac{{\alpha}{\omega}}{\left(\!\,1+
{\alpha}{\omega}\!\right) ^{2}}}} \,t)+
i \sqrt {{{1+\alpha\,\omega}\over{\omega
(1+\alpha\,\omega+\alpha^2\omega^2)}}} \sin(
\omega\sqrt{1-{\frac{{\alpha}{\omega}}{\left(\!\,1+
{\alpha}{\omega}\!\right) ^{2}}}} \,t)\\
\sqrt {{1\over\omega}+\alpha\,\left (\cos(\omega\,t)\right )^{2}}
e^{i \int_0^t{dt\over
{{1\over\omega}+\alpha\,\left (\cos(\omega\,t)\right )^2} }}\\
e^{i{ \pi\over\sqrt{1+\alpha\omega} }}
\left( {1\over\sqrt{\omega}} \cos(\omega
\sqrt{1 - \alpha\omega} \,(t-{\pi\over2\omega}))+
i {1\over\sqrt{\omega(1-\alpha\omega)}}
\sin(\omega\sqrt{1-\alpha\omega} \,(t-{\pi\over2\omega}))\right) \\
\end{array} \right.
\end{equation}
is a solution of the subsidiary classical equation (\ref{eq:sys_equ})
of two-dimensional harmonic oscillator for the same
time-intervals as in (\ref{eq:Omega}).
%
%
%
%
%
The complex-conjugate of $\epsilon^{*}(t)$ is the other linear-independent
solution of (\ref{eq:sys_equ}).
Figure 2 shows the parametric plot of the real and imaginary parts of
$\epsilon(t)$.\ \

As it is shown in  Appendix 1, Wronsky determinant is one {\it classical}
integral of motion for the equation (\ref{eq:sys_equ}).
\begin{equation}\label{eq:wro}
D_W(t)=\epsilon(t)\,\dot{\epsilon}^*(t) -
\dot{\epsilon}(t)\,\epsilon^*(t)= - 2 i .
\end{equation}
We will use this invariant beyond to study the properties of {\it quantum}
integral of motion, especially at the calculation of their commutator.

So we have solved completely the {\it classical} problem of nonstationary
harmonic oscillator with switching the frequency (\ref{eq:Omega}).
Let we consider the {\it quantum} problem of nonstationary harmonic
oscillator with Hamiltonian (\ref{eq:HH}), when the frequency is switching
by the same way as classical oscillator (\ref{eq:Omega}). The quantum
system evolves in time according to the Schr\"{e}dinger equation
\begin{equation}\label{eq:Sch_equ}
i \hbar {\partial\over{\partial t}} \Psi(q,t) = \hat{H}(t) \Psi(q,t),
\end{equation}
where $\Psi(q,t)=<q|\Psi,t>$ is a function of the state
in coordinate representation. We define creation and annihilation operators
( $\Omega_0=\Omega(0)$ )
\begin{equation}\label{eq:boson}
\hat{a}=\left[ {1\over{2\hbar m \Omega_0}}\right] ^{1\over{2}}
(m\Omega_0 \hat{q} + i\,\hat{p})\ \ \ \ \
\hat{a}^{\dagger}=\left[ {1\over{2 \hbar m \Omega_0}}\right]^{1\over{2}}
(m\Omega_0 \hat{q} - i\,\hat{p})
\end{equation}
and it is easy to show that $[a,a^{\dagger}]=1$, because
quantum-mechanical position and  momentum operators $\hat{q}$
and $\hat{p}$ obey the usual commutation relation
$[\hat{q},\hat{p}]=i\hbar$.
According to the general method \cite{trifonov1},\cite{trifonov4} presented
also in \cite{angelow} (see furmula (6) therein), the linear invariant
which corresponds to our particular Hamiltonian has the form;
\begin{equation}\label{eq:inv}
A(t)= {1\over2}\left[
-(\sqrt{\Omega_0} \, \epsilon(t)+
{i\over \sqrt{\Omega_0}} \, \dot{\epsilon}(t) \, )\hat{a}^{\dagger}+
(\sqrt{\Omega_0} \, \epsilon(t)-
{i\over \sqrt{\Omega_0}} \, \dot{\epsilon}(t) \, )\hat{a} \right]
\end{equation}
Operator $A^{\dagger}(t)$ is Hermitian conjugate to $A(t)$ and has
the form;
\begin{equation}\label{eq:inv1}
A^{\dagger}(t)= {1\over2}\left[
(\sqrt{\Omega_0} \, \epsilon^{*}(t)+
{i\over \sqrt{\Omega_0}} \, \dot{\epsilon}^{*}(t) \,)\hat{a}^{\dagger}-
(\sqrt{\Omega_0} \, \epsilon^{*}(t)-
{i\over \sqrt{\Omega_0}} \, \dot{\epsilon}^{*}(t) \, )\hat{a} \right].
\end{equation}
It is also an integral of motion for the quantum system.
Using $[a,a^{\dagger}]=1$ and Wronsky determinant
(\ref{eq:wro}) it is easy to show,
that these invariants obey commutation relation
\begin{equation}\label{eq:com}
[A(t),A^{\dagger}(t)]=1 ,
\end{equation}
i.e. their commutator does not depends on time, either. Thus the operators
$A(t)$ and $A^{\dagger}(t)$ satisfy boson commutation relation in
every moment $t$.
The Hermitian and anty- Hermitian parts of $A(t)$ and $A^{\dagger}(t)$
are also integrals of motion, so that one can construct a new pair
integrals of motion
\begin{equation}\label{eq:int_mot}
\hat{Q}_0(t)=\sqrt{\hbar\over {2 m \Omega_0}}
(\hat{A}^{\dagger}(t)+\hat{A}(t))={1\over\sqrt{\Omega_0}}
({\cal I}m \, (\dot{\epsilon}) \,\, \hat{q} -
{{{\cal I}m \, (\epsilon)}\over m} \,\, \hat{p} ),
\end{equation}
\begin{equation}\label{eq:int_mot1}
\hat{P}_0(t)=i \sqrt{{\hbar m \Omega_0}\over 2}
(\hat{A}^{\dagger}(t)-\hat{A(t)})=\sqrt{\Omega_0}
(-m{\cal R}e \, (\dot{\epsilon}) \,\, \hat{q} +
{{\cal R}e \, (\epsilon)} \,\, \hat{p} ) .
\end{equation}
By direct calculation using the found expression for $\epsilon(t)$
(\ref{eq:epsilon}), and the equality (\ref{eq:sys_equ})
one can check that
${d{\hat Q}_0\over dt}={\partial {\hat Q}_0\over \partial t}-
{i\over \hbar}[{\hat Q}_0,{\hat H}]=0$, respectively
${d{\hat P}_0\over dt}={\partial {\hat P}_0\over \partial t}-
{i\over \hbar}[{\hat P}_0,{\hat H}]=0$.
The physical sense of these two integrals of motion is that the
operators $\hat{Q}_0(t)$ and $\hat{P}_0(t)$ are the initial operators
of coordinate and momentum of the quantum oscillator, scaled with
a factor $\sqrt{\Omega_0}\epsilon(0)$ \,\,\,
( $\dot{\epsilon}(0)=-\dot{\epsilon}^*(0)={i\over\epsilon(0)}
\Rightarrow  {\cal I}m \, (\dot{\epsilon}(0))={1\over\epsilon(0)}$ );
\begin{equation}\label{eq:int_mot2}
\hat{Q}_0(t)= \hat{Q}_0(0)= {1\over{\sqrt{\Omega_0}\epsilon(0)}} \hat{q}
\end{equation}
\begin{equation}\label{eq:int_mot3}
\hat{P}_0(t)= \hat{P}_0(0)= \sqrt{\Omega_0}\epsilon(0) \hat{p} ,
\end{equation}
where the initial condition for $\epsilon(0)$ and $\dot{\epsilon}(0)$
were used ( see (\ref{eq:epsilon}) and (\ref{eq:der_eps}) in Appendix 1,
respectively ).

$\hat{Q}_0(t)$ and $\hat{P}_0(t)$ for an arbitrary quantum system
are related with the initial operators
of coordinate and momentum $\hat{q}(0)$ and $\hat{p}(0)$ in the
time moment $t=0$. The evolutions of $\hat{q}$ and $\hat{p}$ for the
quantum system are compensated in proper way with ${\cal R}e$ and
${\cal I}m$ parts of $\epsilon(t)$ and $\dot{\epsilon}(t)$
to keep $\hat{Q}_0(t)$ and $\hat{P}_0(t)$ constant. The special choice
of operators $A(t)$ and $A^{\dagger}(t)$, which are expressed via solutions
of classical harmonic oscillator (\ref{eq:epsilon}) provides this
property of the quantum integrals of motion in the method of the
linear invariants \cite{trifonov4}. In particular this is true
for the present quantum oscillator whose motion is determined
in terms of the found solution (\ref{eq:epsilon}). Indeed, the quantum mean
value of the integrals of motion $\hat{Q}_0(t)$ and $\hat{P}_0(t)$ are
constants, as we will convince later.

Let us determine the evolution of the first and second moments
of the operators $\hat{q}$ and $\hat{p}$ for the particular oscillator
with frequency (\ref{eq:Omega}).
In \cite{trifonov5} was proved the following theorem;
the necessary an sufficient condition one quantum systems
to preserve an equality in the Schr\"{e}dinger uncertainty
relation is the states of the system to be eigenstates of the
operator $\hat{A}= u\hat{a}+v\hat{a}^{\dagger}+w$. Here $u,v$ and $w$
are arbitrary complex numbers, with one connection; $|u|^2-|v|^2=1$,
and the states are called Schr\"{e}dinger Minimum Uncertainty States.
The invariants (\ref{eq:inv}), (\ref{eq:inv1}) obey this condition
( the proof is based again on the expression of the Wronsky determinant
(\ref{eq:wro}) ).
Let us consider the oscillator in such states, i.e. $|\Psi,t>=|SMUS>$
and $z$ is the corresponding eigenvalue; $\hat{A}|SMUS>=z|SMUS>$.
At this point, there are two ways to solve the problem; to express
$\hat{q}$ and $\hat{p}$ in terms of $\hat{A}$ and $\hat{A}^{\dagger}$,
respectively $u$, $v$ in terms of $\epsilon(t)$, $\dot{\epsilon}(t)$
(\ref{eq:epsilon}) and to find the first and second moments for this
specific Hamiltonian (\ref{eq:HH}). The second way is to use
the method of linear invariants where these moments are obtained
in general form, and to establish the connection with particular
quantum harmonic oscillator with a switching frequency. We chose the
second way, following the idea of the method for one-dimensional case
from \cite{trifonov4,angelow}.

Presenting $\hat{A}$ and $\hat{A}^{\dagger}$ from \cite{angelow}
in terms of $\hat{q}$ and $\hat{p}$
and solving this system about $\hat{q}$ and $\hat{p}$ we receive their
quantum evolutions, expressed by the solutions
of the classical two-dimensional harmonic oscillator
($\epsilon=\epsilon_1+i\epsilon_2$)
\begin{equation}\label{eq:q}
\hat{q}=\sqrt{\hbar a(t)} ( \epsilon(t) \hat{A}^{\dagger}+
\epsilon^*(t)\hat{A})
\end{equation}
\begin{equation}\label{eq:p}
\hat{p}= -\sqrt{{\hbar\over a(t)}}\left [
\left (b\epsilon(t)-{\dot{\epsilon}(t)\over2}-
{1\over4}{\dot{a}(t)\over a(t)}\epsilon(t)\right )\hat{A}^{\dagger} +
\left (b\epsilon^*(t)-{\dot{\epsilon}^*(t)\over2}-
{1\over4}{\dot{a}(t)\over a(t)}\epsilon^*(t)\right )\hat{A} \right ] ,
\end{equation}
where $a(t)$, $b(t)$ and $c(t)$ are the time-dependent coefficients in the
general quadratic Hamiltonian \cite{trifonov4,angelow} in
front of $\hat{p}^2$, $\hat{p}\hat{q}+\hat{q}\hat{p}$ and $\hat{q}^2$.
Taking quantum mean value of $\hat{q}$ and $\hat{p}$ we obtain
the behaviour of their first moments
\begin{equation}\label{eq:<q>}
<\hat{q}>=\sqrt{\hbar a(t)} ( \epsilon(t) z^* + \epsilon^*(t) z)
\end{equation}
\begin{equation}\label{eq:<p>}
<\hat{p}>= -\sqrt{{\hbar\over a(t)}}\left [
\left (b(t)\epsilon(t)-{\dot{\epsilon}(t)\over2}-
{1\over4}{\dot{a}(t)\over a(t)}\epsilon(t)\right ) z^* +
\left (b(t)\epsilon^*(t)-{\dot{\epsilon}^*(t)\over2}-
{1\over4}{\dot{a}(t)\over a(t)}\epsilon^*(t)\right ) z \right ]  .
\end{equation}
For our case (\ref{eq:HH})  the Hamiltonian's coefficients are
$a(t)={1\over2m}$, $b(t)=0$ and $c(t)={m\Omega^2(t)\over2}$.
Consequently, the evolutions of quantum mean value of the coordinate
$<\hat{q}>$ and  the momentum $<\hat{p}>$ has the forms;
\begin{equation}\label{eq:<q>1}
<\hat{q}>=\sqrt{{\hbar\over2m} } ( \epsilon(t) z^* + \epsilon^*(t) z)
\end{equation}
\begin{equation}\label{eq:<p>1}
<\hat{p}>= \sqrt{{\hbar m}\over2}
\left (\dot{\epsilon}(t) z^* + \dot{\epsilon}^*(t) z \right ) .
\end{equation}
Here $z$ is the corresponding eigenvalue of the eigenstates $|SMUS>$, in
whose states the quantum values are obtained.
The phase diagram shown on figure 3 presents the evolution of the
quantum harmonic oscillator with a frequency (\ref{eq:Omega})
( $\hbar=1 J s, \, m=1 g$, $\, \alpha=.5 s$,
$\, \omega=1 Hz$, $\, z=1+i 0.2$ ). The oscillator evolves as an ellipse
when the frequency is a constant ( $t<0$ and $t>{\pi\over2\omega}$).
The ellipse with the smaller horizontal axis corresponds to
the region $t<0$. The bold curve presents the region of the switching
frequency. The point in the first quadrant corresponds to the
quantum mean value of the pair invariants $(<Q_0(t)>,<\hat{P}_0(t)>)$,
(\ref{eq:int_mot}) and (\ref{eq:int_mot1}) respectively.

There are three second independent moments $\sigma_q,\sigma_p$
and $c_{qp}$ ( $c_{pq}=c_{qp}$ ). The quantum deviations
$\sigma_q^2,\sigma_p^2$ as expressions of $\epsilon(t)$ and
$\dot{\epsilon}(t)$ were found for first time in general case of the method
of linear invariants in \cite{trifonov2} ;
\begin{equation}\label{eq:sigma_q}
\sigma_{q}^2(t) = \hbar \,\,a(t) \,|\epsilon(t)|^2,
\end{equation}
\begin{equation}\label{eq:sigma_p}
\sigma_p^2(t) = {\hbar\over a(t)}\left[{1\over{4|\epsilon(t)|^2}}+
\left( b(t)|\epsilon(t)|-{1\over2}{d|\epsilon(t)|\over dt}-
{\dot{a}(t)\over {4a(t)}}|\epsilon(t)|\right)^2\right].
\end{equation}
The third second moment, cofluctuation $c_{qp}$, in terms of $u(t)$ and
$v(t)$ was found in \cite{yuen}, but as an expression of the solution
$\epsilon(t)$ and its first derivative $\dot{\epsilon}(t)$ of the equation
(\ref{eq:sys_equ}) in \cite{angelow};
\begin{equation}\label{eq:c_qp}
c_{qp}^2(t) = \hbar^2 |\epsilon(t)|^2 \left( b(t)|\epsilon(t)|-
{1\over2}{d|\epsilon(t)|\over dt}-
{\dot{a}(t)\over {4a(t)}}|\epsilon(t)|\right)^2 .
\end{equation}
As far as we have $\epsilon(t)$ in explicit form (\ref{eq:epsilon}) and
$a(t)={1\over2m}$, $b(t)=0$
all second moments for the quantum harmonic oscillator with
Hamiltonian (\ref{eq:HH}) and frequency (\ref{eq:Omega}) are determined
completely;
\begin{equation}\label{eq:sigma_q1}
\sigma_{q}^2(t) = {\hbar\over2m} \,|\epsilon(t)|^2,
\end{equation}
\begin{equation}\label{eq:sigma_p1}
\sigma_p^2(t) = {{\hbar m}\over2}\left[{1\over{|\epsilon(t)|^2}}+
\left({d|\epsilon(t)|\over dt} \right)^2\right].
\end{equation}
\begin{equation}\label{eq:c_qp1}
c_{qp}^2(t) = {\hbar^2\over4} |\epsilon(t)|^2
\left({d|\epsilon(t)|\over dt}\right)^2 .
\end{equation}
Using these formulas we have been plotted the evolutions of the
quantum mean values $\sigma_q^2,\sigma_p^2$ and cofluctuation
$c_{qp}^2$, presented on figure 3. We can observe that the quantum
oscillator possess all kind Schr\"{e}dinger minimum uncertainty states
$|SMUS>$; coherent, squeezed and cofluctuant states. The term
{\it subfluctuant} has been suggested by Glauber \cite{glauber}
as more accurate than {\it squeezed}, but we will use the traditional
{\it squeezed} for the states with a subfluctuation below the vacuum in
a coherent state and
{\it cofluctuant} for the states with nonzero cofluctuation.
For example the oscillator is in a squeezed-cofluctuant
state in the region $t<0$, as $c_{qp}\neq 0$. They are ideal squeezed
\cite{caves1,caves2} only when the cofluctuation $c_{qp}=0$. In this
situation the Schr\"{e}dinger uncertainty relation devolves in Heisenberg one.
In the region of the switching frequency the state is squeezed-cofluctuant,
too. The same situation exist in the region $t>{\pi\over2\omega}$ except
in the time moments
\begin{equation}\label{eq:tim_mom}
t_n ={\pi\over2\omega}+
\left ({1\over2}+n\right )
{\pi\over{4\omega\sqrt{1-{\alpha\omega\over(1+\alpha\omega)^2}}}} ,
\end{equation}
($n=1,2,3,...$) were the state is coherent ( no squeezing, no
cofluctuations - dimensionless fluctuations are equal; \,\,
$m\Omega(t_n)\sigma_q^2=
{\sigma_p^2/m\Omega(t_n)}= {\hbar /2}$,
$z$ is equals to Glauber's complex number $z=\alpha$ ).
The analysis shows that in the moment $t=0$ for the initial conditions
$\epsilon(0)=\sqrt{{1+\alpha\omega}\over\omega}$ and
$\dot{\epsilon}(0)=i\sqrt{\omega\over{1+\alpha\omega}}$ the oscillator is
in the ideal squeezed state.

An other possible description of the quantum problem is based on the
important Wigner function, which we are going to concern here.
The all information on the quantum system is contained in the time-dependent
density operator $\hat{\rho}(t)$, satisfying the conditions of hermiticity,
normalization and nonnegativity.\ \

For pure states, the density operator which is projector on the
state $|\Psi,t>$, i.e.
\begin{equation}\label{eq:den_ope}
\hat{\rho}(t)=|\Psi,t><t,\Psi|
\end{equation}
satisfies the extra condition
$$ \hat{\rho}^2(t)=\hat{\rho}(t) ,$$
and consequently,
\begin{equation}\label{eq:trace}
Tr \hat{\rho}^2(t)=1 .
\end{equation}
The density operator (\ref{eq:den_ope}) obeys the {\it equation
for invariants} ${d\hat{\rho}(t)\over dt}=0$ ( which differs from
Heisenberg equation of motion, see for example \cite{angelow} )
\begin{equation}\label{eq:inv_equ}
{\partial\hat{\rho}(t)\over \partial t}+{1\over i\hbar}
[\hat{\rho}(t),\hat{H}(t)]=0
\end{equation}
Instead of density matrix one can consider Wigner function \cite{wigner}.
Such formulation of the quantum problem by means of functions on phase space
is very convenient for quantum systems having classical analog.
In Weil representation mean value of every physical variable is an integral
of this variable with the distribution function over the all phase space.
This procedure is a full analog to the classical one.

The Wigner function is Fourie-transformation of the coordinate
representation of density operator.
\begin{equation}\label{eq:wigner}
W(q,p)=\int_{\infty}^{\infty}{\rho(q+{v\over2},q-{v\over 2}) exp(-i p v)}dv
\end{equation}
where $\rho(q,q',t)=<q|\hat{\rho}(t)|q'>$ is the density matrix.

With the help of (\ref{eq:den_ope}), one can calculate the Wigner function
(\ref{eq:wigner}) in states $|SMUS>$;
\begin{equation}\label{eq:wigner2}
W(q,p,\epsilon(t),{d|\epsilon(t)|\over dt})=
\end{equation}
$$ {2\over \pi\hbar} exp \left \{ {-{2\over{\hbar}^2}}
\left [ \sigma_q^2 (p-<\hat{p}>)^2 - 2 \sigma_{qp}(p-<\hat{p}>)(q-<\hat{q}>)+
\sigma_p^2 (q-<\hat{q}>)^2 \right ] \right \} ,$$
where for the determinant of the variances matrix $det (\sigma(t))$
we have used the following expression;
\begin{equation}\label{eq:det1}
det (\sigma(t))=\sigma_{p}^2\sigma_{q}^2-c_{qp}^2={\hbar^2\over 4}.
\end{equation}
As far as $\sigma_{p},\sigma_{q}$ and $c_{qp}$
(\ref{eq:sigma_q1}-\ref{eq:c_qp1}) are functions of the
solution of the equation (\ref{eq:sys_equ}), we have obtained
the evolution of Wigner function in terms of $\epsilon(t)$ and
${d|\epsilon(t)|\over dt}$ for our particular case of
frequency (\ref{eq:Omega}). On the figure 5 Wigner function in the time
moment $t=0$ is shown. It corresponds to the initial point of the bold
curve on figure 3, where the phase-space diagram of the quantum harmonic
oscillator is presented.

In conclusion, an explicit solution of the equation for
the classical harmonic oscillator, with smooth switching frequency
has been found.
A detailed analysis of a quantum harmonic oscillator with such frequency
has been done on the base of the method of linear invariants.
It has been shown that such oscillator also possesses cofluctuant states,
different from widely studied Glauber coherent and "ideal" squeezed
states. It has been found the evolution of the Wigner function of such
quantum oscillator with a switching frequency.\ \

\ \

The author is grateful to D.Trifonov for valuable discussions.
This work was partly supported by Bulgarian Scientific Foundation,
grant number F-559.\ \

\section{Appendix 1}

Here we will calculate the Wronsky determinant for the classical equation
of the two-dimensional oscillator (\ref{eq:sys_equ}).
To derive the Wronskian we need to know the first derivative of
the function $\epsilon(t)$ in the three regions;
\begin{equation}\label{eq:der_eps}
{d\epsilon(t)\over dt}= \left \{
\begin{array}{lr}
-\sqrt {\omega(1+\alpha\,\omega+\alpha^2\,\omega^2)\over{1+\alpha\,\omega}}
\sin(\omega\sqrt{1-{\frac{{\alpha}{\omega}}
{\left(\!\,1+{\alpha}{\omega}\!\right) ^{2}}}} \,t)+
i\sqrt {\omega\over{1+\alpha\,\omega}}
\cos(\omega\sqrt{1-{\frac{{\alpha}{\omega}}{\left(\!\,1+
{\alpha}{\omega}\!\right) ^{2}}}} \,t)\\
{e^{i \int_0^t{dt\over
{{1\over\omega}+\alpha\,\left (\cos(\omega\,t)\right )^2} }}\over
\sqrt {{1\over\omega}+\alpha\,\left (\cos(\omega\,t)\right )^{2}}}
{(-\alpha\,\omega\,\sin(2\omega\,t) + i)}\\
e^{i{\pi\over\sqrt{1+\alpha\omega}}}
\left(-\sqrt{\omega(1-\alpha\omega)}
\sin(\omega\sqrt{1-\alpha\omega}\,(t-{\pi\over2\omega}))+
i\sqrt{\omega}
\cos(\omega\sqrt{1-\alpha\omega}\,(t-{\pi\over2\omega}))\right)\\
\end{array} \right.
\end{equation}
By direct calculation using (\ref{eq:epsilon}) and (\ref{eq:der_eps})
we receive the Wronsky determinant;
\begin{equation}\label{eq:wro1}
D_W(t)=\left | \matrix{
\epsilon(t)             & \epsilon^*(t)\cr
{d\epsilon(t)\over dt}  & {d\epsilon^*(t)\over dt}\
\cr}\right |
=\epsilon(t)\dot{\epsilon}^*(t)-\epsilon^*(t)\dot{\epsilon}(t)=-2i .
\end{equation}
Obviously $D_W(\epsilon(t),\dot{\epsilon}(t))$ is an ( classical )
integral of motion for equation (\ref{eq:sys_equ}).\ \

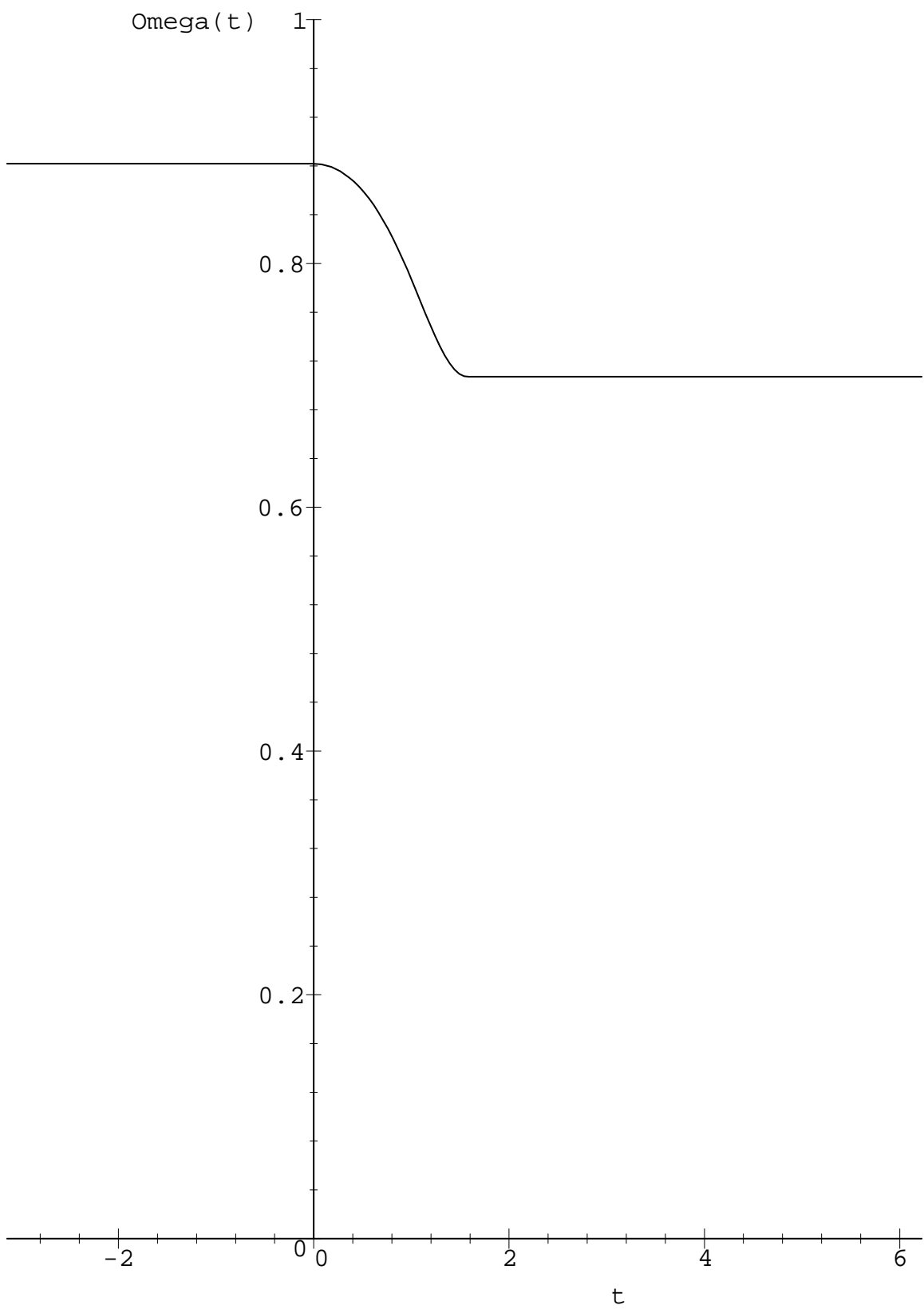
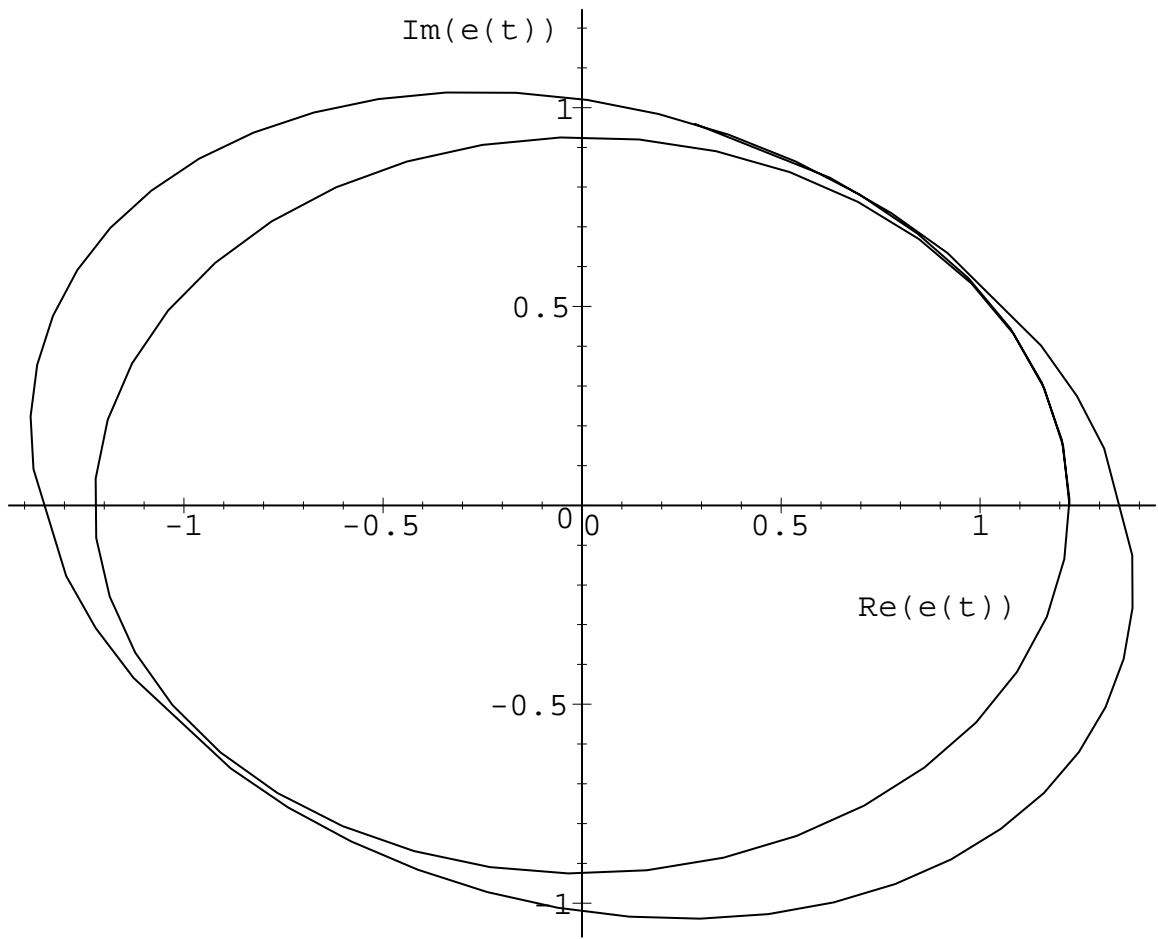
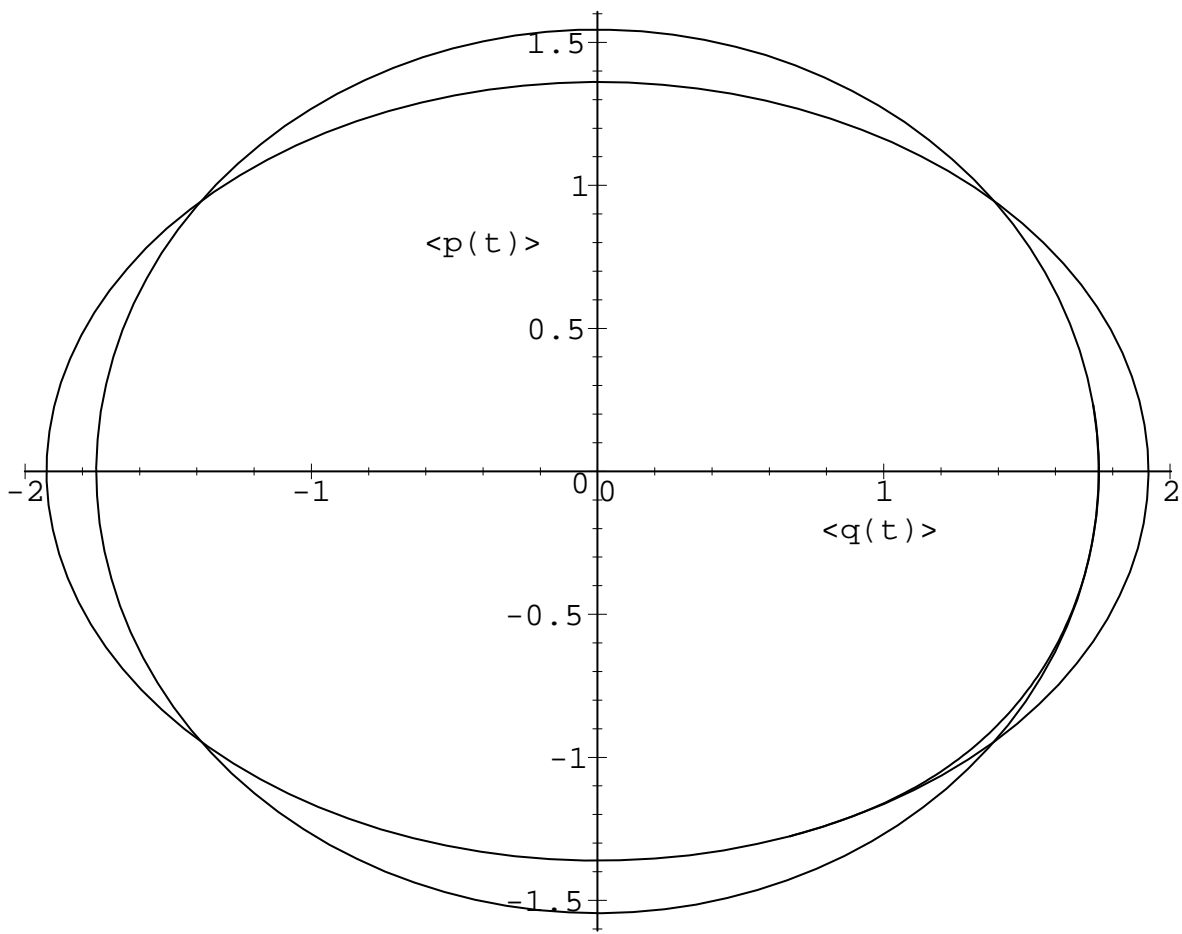
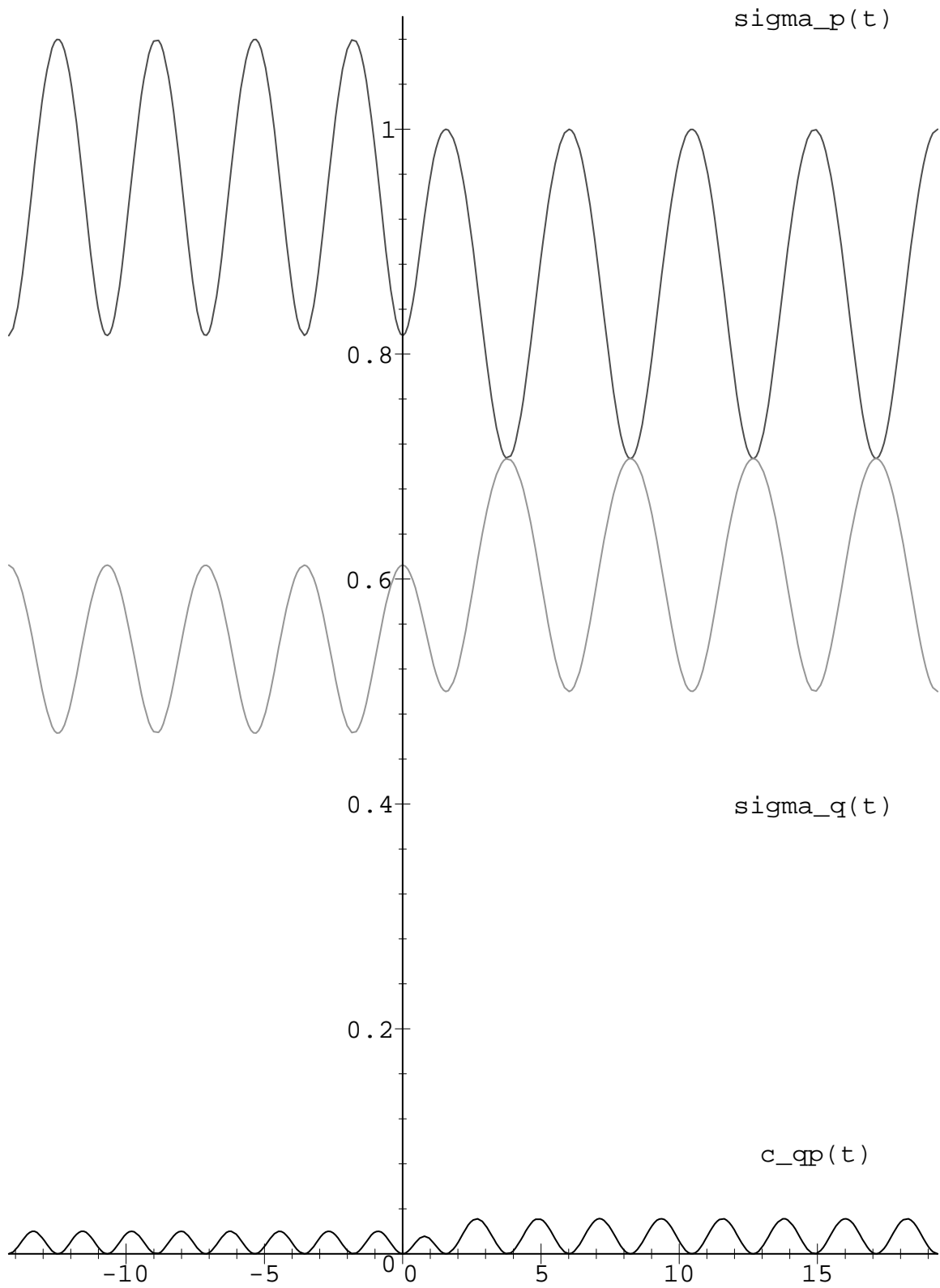
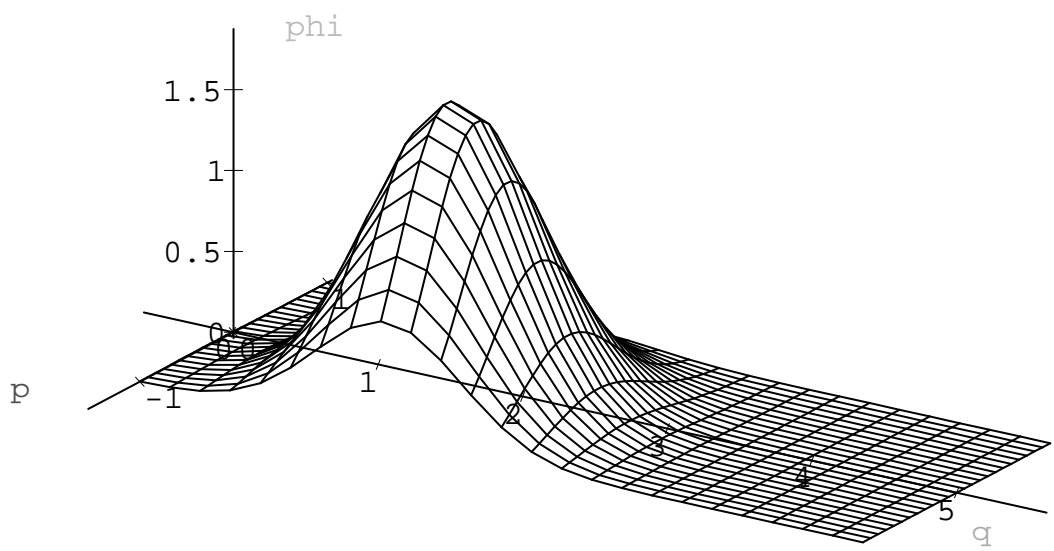
\end{document}